\documentclass[10pt,aps,prl,twocolumn,groupedaddress,showkeys]{revtex4-1}
\usepackage{amsmath}
\usepackage{graphicx}
\DeclareGraphicsExtensions{.pdf,.eps,.png,.jpg,.mps}
\usepackage{threeparttable}
\usepackage{color}

\begin{document}

\title{High-responsivity graphene photodetectors integrated on silicon microring resonators}

\author{Simone Schuler$^{1,2, \ddagger}$}
\email[E-mail: ]{simone.schuler@tuwien.ac.at}
\author{Jakob E. Muench$^{2, \ddagger}$}
\author{Alfonso Ruocco$^2$, Osman Balci$^2$,  Dries van Thourhout$^3$, Vito Sorianello$^4$, Marco Romagnoli$^4$, Kenji Watanabe$^5$, Takashi Taniguchi$^5$, Ilya Goykhman$^{2,6}$, Andrea C. Ferrari$^2$}
\author{Thomas Mueller$^1$}
\email[E-mail: ]{thomas.mueller@tuwien.ac.at}

\affiliation{$^1$Vienna University of Technology, Institute of Photonics, Gußhausstraße 27-29, 1040 Vienna, Austria}
\affiliation{$^2$Cambridge Graphene Centre, University of Cambridge, Cambridge CB3 0FA, UK}
\affiliation{$^3$Ghent University-IMEC, Photonics Research Group, Technologiepark-Zwijnaarde 126, 9052 Gent, Belgium}
\affiliation{$^4$Consorzio Nazionale per le Telecomunicazioni, Photonic Networks and Technologies National Lab, via Moruzzi 1, 56124 Pisa, Italy}
\affiliation{$^5$National Institute for Materials Science, 1-1 Namiki, Tsukuba, 305-0044 Japan}
\affiliation{$^6$Technion - Israel Institute of Technology, Haifa, 3200003, Israel}
\affiliation{$^\ddagger$These authors contributed equally to this work.}

\begin{abstract}
Graphene integrated photonics provides several advantages over conventional Si photonics. Single layer graphene (SLG) enables fast, broadband, and energy-efficient electro-optic modulators, optical switches and photodetectors (GPDs), and is compatible with any optical waveguide. The last major barrier to SLG-based optical receivers lies in the low responsivity - electrical output per optical input - of GPDs compared to conventional PDs. Here we overcome this shortfall by integrating a photo-thermoelectric GPD with a Si microring resonator. Under critical coupling, we achieve $>$90\% light absorption in a  $\sim$6 $\mu$m SLG channel along the Si waveguide. Exploiting the cavity-enhanced light-matter interaction, causing carriers in SLG to reach  $\sim$400 K for an input power of  $\sim$0.6 mW,  we get a voltage responsivity  $\sim$90 V/W, demonstrating the feasibility of our approach. Our device is capable of detecting data rates up to 20 Gbit/s, with a receiver sensitivity enabling it to operate at a 10$^{-9}$ bit-error rate, on par with mature semiconductor technology. The natural generation of a voltage rather than a current, removes the need for transimpedance amplification, with a reduction of the energy-per-bit cost and foot-print, when compared to a traditional semiconductor-based receiver.
\end{abstract}

\maketitle
\section{Introduction}
The same-chip integration  \cite{Atabaki2018Integrating} of active and passive optical components with electronics offers a cost- and energy-efficient solution for short- and long-reach optical interconnects  \cite{cheng2018recent,subbaraman2015recent}. Single layer graphene (SLG) is an ideal material for integrated photonics  \cite{Romagnoli2018Graphene,Bonaccorso2010Graphene}, promising e.g. high-speed (potentially $>$200 GHz)  \cite{urich2011intrinsic,Xia2009ultrafast} and broadband (ultraviolet to far-infrared)  \cite{Koppens2014Photodetectors} operation that could lift bandwidth (BW) ($\sim100$ GHz)  \cite{vivien2012zero,Salamin2018GHz} and spectral ($<1600$ nm) \cite{Michel2010High} limitations of existing technologies, such as Ge/Si \cite{thomson2016roadmap,chrostowski2015silicon} and InGaAsP/InP \cite{arafin2018advanced,Nagarajan2010InP}. A variety of waveguide (WG)-integrated SLG-based photonic devices have been reported \cite{Liu2011graphene,phare2015graphene,giambra2019high,Sorianello2018graphene,Cassese2017Capacative,Pospischil2013CMOS,Gan2013chip,Wang2013high,Shiue2015High,Schuler2016Controlled,schuler2018graphene,Goykhman2016On,Muench2019Waveguide,Schall2018record,Ma2018Plasmonically,ding2020ultra}, including electro-absorption (EAMs) \cite{Liu2011graphene,phare2015graphene,giambra2019high} and electro-refraction modulators (ERMs) \cite{Sorianello2018graphene}, optical switches \cite{Romagnoli2018Graphene,Cassese2017Capacative} and photodetectors (GPDs) \cite{Pospischil2013CMOS,Gan2013chip,Wang2013high,Shiue2015High,Schuler2016Controlled,schuler2018graphene,Goykhman2016On,Muench2019Waveguide,Schall2018record,Ma2018Plasmonically,ding2020ultra}. SLG and layered materials can be integrated with passive Si photonic WGs \cite{Pospischil2013CMOS,Gan2013chip,Wang2013high,Shiue2015High,Schuler2016Controlled,schuler2018graphene} or any other passive WG technology \cite{Romagnoli2018Graphene}, including Si$_3$N$_4$ \cite{Muench2019Waveguide,gao2018high}, sapphire \cite{cheng2015graphene}, Ge \cite{wang2019design}, and polymers \cite{kim2012graphene,kleinert2016graphene}, extending the spectral range and scope of possible applications \cite{wang2019design,qu2018waveguide}.

SLG's optical absorption is $\sim$2.3\% under normal incidence \cite{Nair2008Fine}, which limits the photoresponse in top-illuminated GPDs \cite{Koppens2014Photodetectors}. This absorption can be increased in a WG configuration through the interaction with the evanescent field of the optical WG mode \cite{Youngblood2016integration}. However, since the mode-overlap with SLG's monatomic cross-section typically restricts absorption to $\sim0.01-0.1$ dB/$\mu$m ($\sim0.2-2$\%/$\mu$m)  \cite{Romagnoli2018Graphene}, device lengths $\sim$100 $\mu$m are needed for near-complete ($>$90\%) light absorption, with adverse effects on foot-print and capacitance that scale with device size \cite{Xia2009ultrafast}. The resulting trade-off between length and absorption has implications for GPDs that operate via the photo-thermoelectric effect (PTE) \cite{gabor2011hot,Song2011Hot,Xu2010Photo}: Due to slow($\sim$ps \cite{Brida2013ultrafast}) heat dissipation to the lattice via phonon mediated cooling \cite{Brida2013ultrafast,tomadin2013nonequilibrium}, photo-excitation leads to the formation of a hot-carrier distribution in SLG \cite{gabor2011hot,Song2011Hot,Koppens2014Photodetectors}. The associated electron temperature $T_e$ can be substantially above the lattice temperature $T_0$ \cite{Soavi2018Broadband} and leads to a photovoltage according to \cite{gabor2011hot,Song2011Hot,Xu2010Photo}:
\begin{equation}\label{eq:V-PTE}
V_\text{PTE}=\int S(x) \cdot\nabla T_e(x)\mathrm{d}x
\end{equation}
if both a $T_e$ gradient and a spatially varying Seebeck coefficient $S$ (controlled by the chemical potential $\mu_c$) are present \cite{gabor2011hot}. In order to achieve a high ($>$mV) $V_\text{PTE}$, it is better to absorb the incident electromagnetic energy over small ($<10$ $\mu$m) length scales, leading to localised electronic heating for a higher ( $\sim$tens K) $T_e$ and $\nabla T_e(x)$.

Different approaches have been explored to increase and confine light absorption in free-space-coupled \cite{Echtermeyer2011Strong,Engel2012Light,Furchi2012Microcavity} and WG-integrated GPDs \cite{Schuler2016Controlled,Goykhman2016On,schuler2018graphene,Muench2019Waveguide}, e.g. by embedding SLG into optical cavities \cite{Engel2012Light,Furchi2012Microcavity}, slot WGs \cite{Schuler2016Controlled}, plasmonic structures \cite{Echtermeyer2011Strong,Goykhman2016On,Muench2019Waveguide,Ma2018Plasmonically}, or by enhancing the light-matter interaction using sub-wavelength structures \cite{schuler2018graphene} but these have coupling and propagation losses \cite{Ma2018Plasmonically}, limitation in field enhancement \cite{Shiue2015High,Schall2018record}, or carrier mobility $\mu$ \cite{Schuler2016Controlled,ding2020ultra}, fabrication flows incompatible with complementary metal-oxide-semiconductor (CMOS) processing \cite{Ma2018Plasmonically,Shiue2015High,ding2020ultra}, bias-induced dark currents \cite{Ma2018Plasmonically,Schall2018record}, or a combination thereof \cite{Ma2018Plasmonically,Shiue2015High,ding2020ultra}. Thus, the demonstration of GPDs on photonic integrated circuits (PICs) that leverage SLG's unique hot-carrier dynamics and maximise the voltage responsivity $R_\text{[V/W]}=V_\text{PTE}/P_\text{in}$, with $P_\text{in}$ the incident optical power in the WG, is challenging.

Here, we present GPDs integrated on looped WGs, known as microring resonators \cite{bogaerts2012silicon}, which act as PIC-embedded resonant cavities. The conversion of the incident light into an electrical signal occurs via the PTE effect. The GPDs directly generates a voltage, which allows us to operate them without bias and dark currents, limiting the GPD noise to thermal Johnson noise, i.e. the noise due to fluctuations of the carrier density \cite{fang2019accurate}. This removes the need of transimpedance amplifiers (TIA) in the read-out electronics, with a reduction of the energy-per-bit cost and system foot-print. With $R_\text{[V/W]}\sim$ 90 V/W, our GPDs pave the way towards SLG integration on Si photonic receivers, overcoming the limitation of photocurrent ($I_\text{ph}$) generating GPDs with responsivities ($R_\text{[A/W]}=I_\text{ph}/P_\text{in}$) lower than that of mature Ge PDs  \cite{Ma2018Plasmonically,Shiue2015High,Schall2018record}. We attribute this to our high ($>10^4\text{ cm}^2/\text{Vs}$) SLG $\mu$ and the combination with the Si microring resonator giving a $\sim$10-fold enhancement of the electric field strength and $>$90\% light absorption in only $\sim$6 $\mu$m SLG on the Si WG.
\section{Results and discussion}
\begin{figure*}
\centerline{\includegraphics[width=180mm]{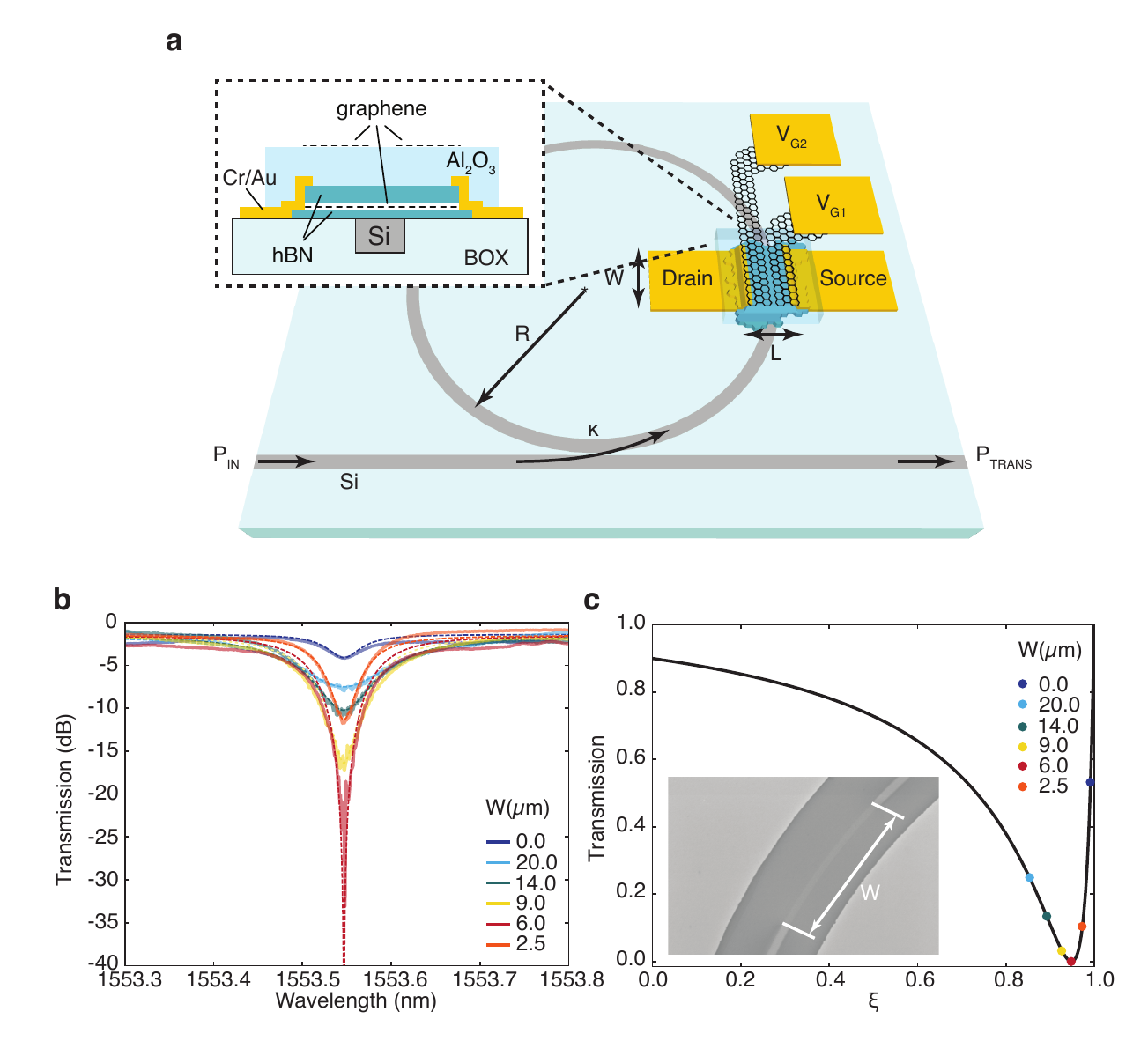}}
\caption{Si ring resonator integrated GPD. (a) Sketch of the device. (b) Transmission of a ring cavity with SLG on top. (c) Transmission at resonance ($\sim$1553.55 nm). The black line is the calculated transmission for $\kappa$ = 10\%. The coloured dots mark the transmission coefficient for various loss coefficients}
\label{fig:schematic}
\end{figure*}
Fig.\ref{fig:schematic} is a scheme of our GPDs, which comprise a layered materials heterostructure (LHM) of SLG and hexagonal boron nitride (hBN). To increase the generated $V_\text{PTE}$ upon optical illumination according to Eq.\ref{eq:V-PTE}, encapsulation of the SLG channel in hBN ensures a high ($>10^4\text{ cm}^2/\text{Vs}$) $\mu$ \cite{purdie2018cleaning} for large ($\sim200$ $\mu$V/K) peak $S$ \cite{Muench2019Waveguide}, according to Mott's formula \cite{gabor2011hot,Song2011Hot}:
\begin{equation}\label{eq:Mott}
S = -\frac{\pi^2 k_\text{B}^2 T_e}{3 e \sigma(\mu,\mu_c)}\frac{\mathrm{d}\sigma(\mu,\mu_c)}{\mathrm{d}\mu_c}
\end{equation}
where $k_\text{B}$ is the Boltzmann constant, $e$ the electron charge, $\sigma=n\mu e$ the conductivity, and $n$ the carrier concentration. Dual-gate SLG electrodes, separated from the LMH by a Al$_2$O$_3$ layer, are employed to tune $S$ in adjacent regions of the device \cite{Schuler2016Controlled}. The LMH is contacted on opposite sides and centrally aligned to the WG of a microring resonator fabricated on a Si-on-insulator (SOI) wafer. The resonator serves a two-fold purpose. First, the higher (compared to the bus WG) intra-cavity energy density \cite{bogaerts2012silicon} results in $\sim$10-fold enhanced light-matter interaction \cite{rabus2007integrated} and can enable near-complete light absorption in the SLG channel if its coverage of the resonator is optimised. Second, the wavelength $\lambda$ selectivity of the resonator \cite{Yariv2002critical,rabus2007integrated}, makes the GPD suitable for wavelength division multiplexing (WDM) \cite{chrostowski2015silicon}, whereby the data rate of a single optical channel is increased by combining signals of different $\lambda$ at the transmitter and separating them at the receiver \cite{Xu2006Cascaded,Zheng2010tunable}.

To find the SLG length $W$ over the ring that enables maximum absorption inside the resonator, we perform an initial experiment on a reference microring cavity with identical parameters (WG thickness $t_\text{WG}$ = 220 nm, WG width $w_\text{WG}$ = 480 nm, ring radius $R$ = 40 $\mu$m). Coupling to the resonator occurs through a 200 nm gap via a single bus WG, Fig.\ref{fig:schematic}a, which has grating couplers on either end, with coupling efficiency \cite{chrostowski2015silicon} $\eta=P_\text{in}/P_\text{fibre}\sim$ 0.28, with $P_\text{fibre}$ the optical power in the fibre connecting source and SOI chip, determined from transmission measurements on reference WGs on the same chip. The power coupling between these two structures depends on the coupling ($\kappa$) and transmission coefficient ($t$), i.e. the scattering matrix elements relating incoming and outgoing electric fields from the coupling region \cite{Yariv2002critical}. These are related via \cite{rabus2007integrated} $\kappa=\sqrt{1-t^2}$. The choice of $\kappa$ affects both the optical (frequency) bandwidth $\delta\nu$  and quality ($Q$) factor ($Q=\nu_0/\delta\nu$, with $\nu_0$ the resonance frequency  \cite{paschotta2008encyclopedia}) of the resonator, resulting in a trade-off between achievable extinction ratio (ER, defined as the ratio of minimum (at resonance) and maximum transmitted optical power  \cite{bogaerts2012silicon}), thus $R_\text{[V/W]}$, and BW. We select $\kappa=10$\%, which allows $BW>10$ GHz in our design \cite{rabus2007integrated}, sufficient for applications in data centre optical interconnects \cite{cheng2018recent}.

The wavelength-dependent transmitted power $P_\text{trans}$ of a ring resonator can be written as \cite{bogaerts2012silicon}:
\begin{equation}\label{eq:transPower}
P_\text{trans} = P_\text{in}\frac{(1-\kappa)+\xi-2\sqrt{(1-\kappa)\xi}\cos(\theta)}{1+(1-\kappa)\xi-2\sqrt{(1-\kappa)\xi}\cos(\theta)}
\end{equation}
where \mbox{$\theta=4\pi^2 R n_\text{eff}/\lambda$} is the round-trip phase shift of the circulating mode and $n_\text{eff}$ is the effective mode index ($n_\text{eff}=\beta/k_0$ with $\beta$ the propagation constant of the mode, defined as the wavevector component along the WG, and $k_0$ the free-space wavevector \cite{reed2004silicon}). Omitting the negligible losses caused by coupling between bus and ring, the term:
\begin{equation}\label{eq:roundtripLoss}
\xi = e^{-W\alpha_\text{SLG}}e^{-2\pi R \alpha_\text{WG}}
\end{equation}
describes the round-trip propagation loss in the ring, with $\alpha_\text{SLG}$ and $\alpha_\text{WG}$, in dB/$\mu$m, the power attenuation coefficients in SLG and Si WG, respectively. When $\theta$=2$\pi m$ ($m$ = 1,2,3\dots), the light in the ring constructively interferes with itself and the cavity is in resonance \cite{bogaerts2012silicon}. From Eq.\ref{eq:transPower}, the transmission drops to zero if $\xi=1-\kappa$. Under this so-called critical coupling \cite{Yariv2002critical}, maximum field enhancement is achieved inside the resonator as the transmitted power goes to zero. With all other parameters fixed in Eq.\ref{eq:roundtripLoss}, changing the SLG-induced losses by changing the coverage length $W$ can therefore be used to tune $\xi$ and achieve critical coupling.
\begin{figure*}
\centerline{\includegraphics[width=160mm]{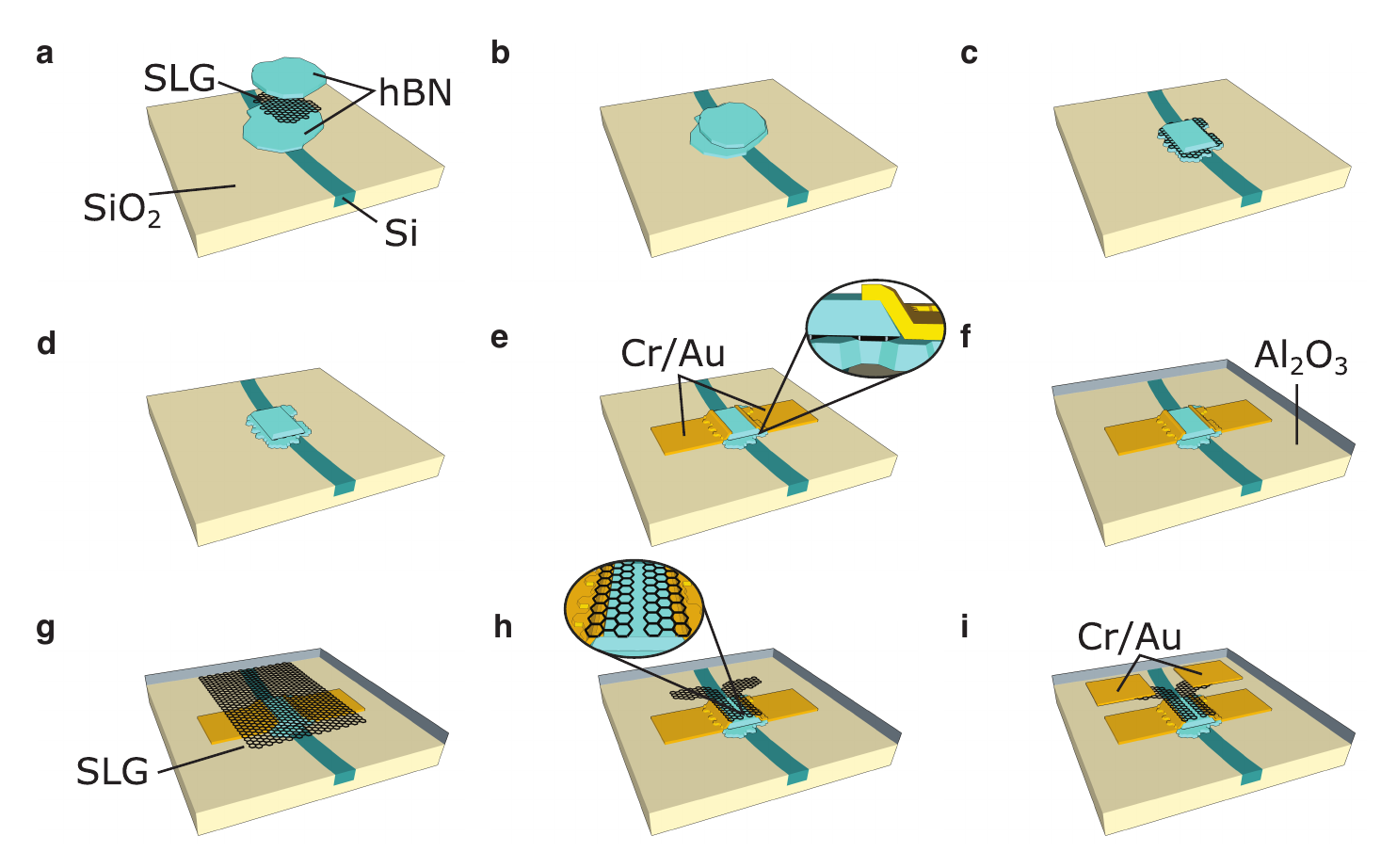}}
\caption{(a) Assembly of hBN/SLG/hBN LMH.(b) Stack placement on photonic circuit and interface cleaning. (c) hBN etching in SF$_6$  plasma. (d) SLG etching in O$_2$  plasma to define channel geometry. (e) Metallization (Cr/Au) for drain-source contacts. (f) Al seed layer evaporation + ALD of Al$_2$O$_3$. (g) Wet transfer of CVD SLG. (h) Split-gate fabrication. (i) Metallization (Cr/Au) for gate contacts.}
\label{fig:fabrication}
\end{figure*}

To find this optimum $W$, we first measure the transmission of an unloaded (no SLG, i.e. W = 0) resonator by coupling light (continuous-wave (CW), TE-polarised) from a tunable laser (Newport TLB6700) into the bus WG, using an optical single-mode fibre, and measuring the transmitted power at the second grating coupler as a function of $\lambda$ around one of the resonance peaks. The results, after calibration for the coupling losses, are shown by the dark-blue line and symbols in Figs.\ref{fig:schematic}b,c, respectively. The microring resonator is not critically coupled at resonance for $\lambda\sim1553.55$ $\mu$m as $P_\text{trans}$ does not vanish, but only part of the incident power is dissipated in the WG. From Eqs.\ref{eq:transPower},\ref{eq:roundtripLoss} we get $\alpha_\text{WG}\sim$1.4 dB/cm.

We then proceed to study the effect of SLG with varying $W$ on the power dissipated in the resonator. We first place a $W$ = 20 $\mu$m long SLG flake, prepared by micro-mechanical cleavage (MC) \cite{Novoselov2005Two} of bulk graphite, transferred using a micro-manipulator and a stamp consisting of polycarbonate (PC) and polydimethylsiloxane (PDMS), and cleaned by immersion in chloroform, over the ring and measure the transmission as before. Using successive electron beam lithography (EBL, Raith e-LINE) steps to define a poly(methyl methacrylate) (PMMA) etch mask and reactive ion etching in O$_2$ to remove excess material, we then reduce $W$ further in several steps down to $2.5$ $\mu$m with transmission measurements in between. The results, Figs.\ref{fig:schematic}b,c, show an initial transmission decrease at resonance with decreasing $W$ before the trend is inverted as $W$ tends to zero. The minimum transmission, indicating critical coupling, is for $W$ = 6 $\mu$m. From Eq.\ref{eq:roundtripLoss} we extract $\alpha_\text{G}\sim$ 0.07 dB/$\mu$m, in agreement with measured \cite{Liu2011graphene} and simulated \cite{Romagnoli2018Graphene} values from literature. Using these values, and the comparison of the transmission curves for $W$ = 6 $\mu$m and $W$ = 0, we estimate the fraction of absorbed light in the SLG channel to be $\sim$92\% under critical coupling.
\begin{figure*}
\centerline{\includegraphics[width=150mm]{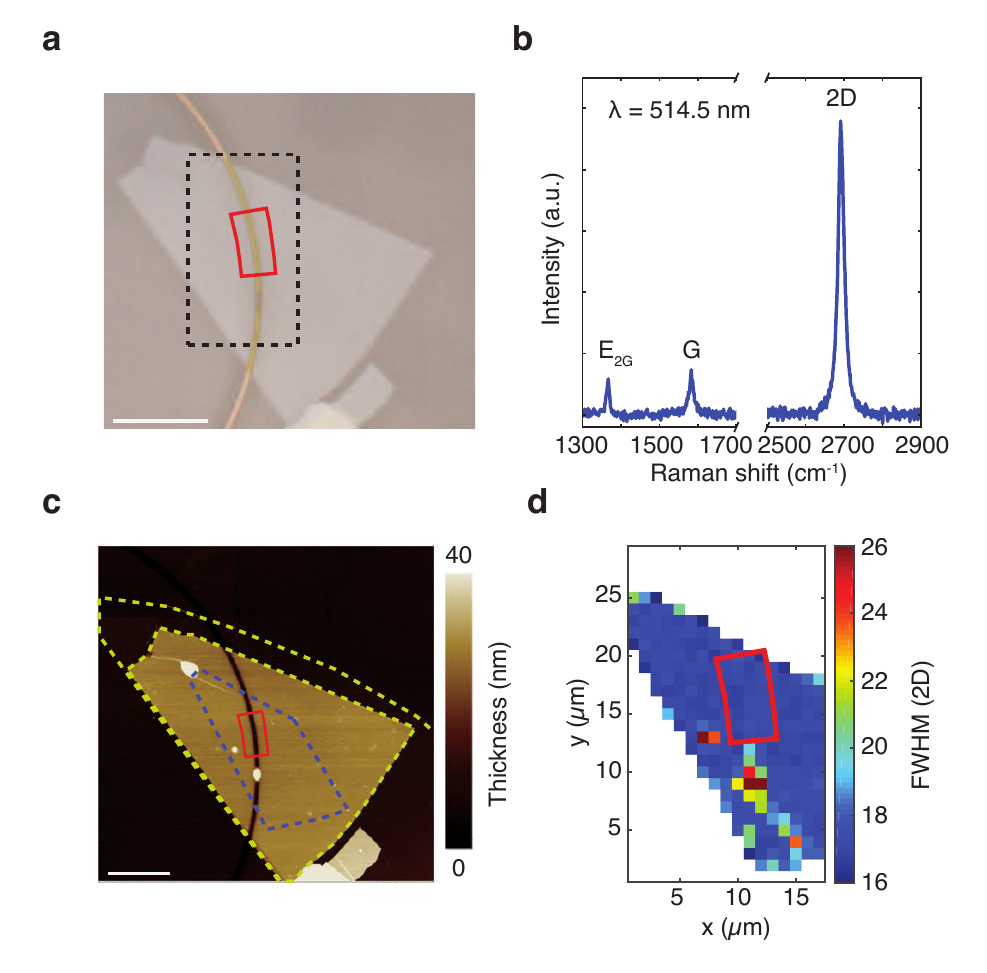}}
\caption{(a) Microscope image of a hBN/SLG/hBN LMH on ring resonator. The black dashed line indicates the area over which the Raman map in (d) is measured. Scale bar, 10 $\mu$m. (b) Raman spectrum measured at the position of the final device.(c) AFM image of the LMH. The yellow dashed line indicates the area of top and bottom hBN. The blue line indicates the SLG area. Scale bar, 10 $\mu$m. (d) Raman map of FWHM(2D). The red box marks the position of the final device in (a,c,d).}
\label{fig:material}
\end{figure*}

Based on these findings, we fabricate the GPD in Fig.\ref{fig:schematic}a with $W$ = 6 $\mu$m from a LMH (hBN encapsulated SLG as channel layer) on top of the ring resonator. This is a prepared as follows \cite{purdie2018cleaning}: SLG and hBN flakes of different thicknesses ($t_\text{bottom}\sim3$ nm, $t_\text{top}\sim20$ nm) are prepared on Si/SiO$_2$ ($t_{\text{SiO}_2}$ = 285 nm) by MC of bulk graphite (Graphenium) and hBN single crystals grown at high pressure and temperature as detailed in Ref. \cite{TANIGUCHI2007Systhesis}. The thickness of the bottom hBN, $t_\text{bottom}$, is chosen with the following trade-off: sufficiently thin ($<$5 nm) to ensure $\alpha_\text{SLG}$ (thus $\xi$) comparable to the initial experiment used to find $W$, but sufficiently thick ($\sim$ nm) that high ($>10^4 \text{ cm}^2/\text{Vs}$) $\mu$ is achieved, due to reduced (in comparison to SiO$_2$) carrier inhomogenities, roughness, and charge impurities \cite{dean2010boron,Wang2013One,purdie2018cleaning}. A micro-manipulator and a PC/PDMS stamp are then used to pick up and stack the flakes at 50$^{\circ}$C (Fig.\ref{fig:fabrication}a). In order the clean the LMH interfaces, the target photonic chip is then heated to 180$^{\circ}$C \cite{purdie2018cleaning}, while we align the LMH to the Si WG. We then laminate the PC film onto the target substrate, pushing contamination blisters, formed at the SLG/hBN interfaces, out of the GPD channel region and placing the LMH on the WG, Fig.\ref{fig:fabrication}b.

After the PC film is dissolved in chloroform, we perform EBL (Raith EBPG 5200) to define the GPD channel geometry via a PMMA etch mask. To transfer this pattern to the LMH, we use two dry etching steps. First, to achieve etch selectivity between LMH and underlying photonic circuit, we use a reactive ion etcher (RIE, Plasma-Therm) with a forward radio frequency power $\sim$80 W and an SF$_{6}$ flow $\sim$80 sccm. As reported by Ref. \cite{jessen2019lithographic}, these conditions allow fast ($>$200 nm/min) etching of hBN, slow ($<$7 nm/min) etching of SiO$_2$, while SLG is not etched, serving as etch stop on the bottom hBN flake (Fig.\ref{fig:fabrication}c). We then expose the LMH to low power (3W) oxygen (O$_2$) plasma to remove all excess SLG, leaving behind the fully shaped GPD channel (Fig.\ref{fig:fabrication}d). A second EBL step, electron beam evaporation (5 nm Cr/50 nm Au), and lift-off in acetone are then used to contact the exposed SLG channel edges (Fig.\ref{fig:fabrication}e). To fabricate the split-gate structure on top of the LMH, required to create a p-n juntion in the GPD channel, a transparent (at $\lambda \sim 1.55$ $\mu$m) conductor that does not affect $\xi$ (thus $W$) is required. A second layer of SLG, sufficiently high ( $\sim$ tens of nm) above the WG to leave $\alpha_\text{SLG}$ unaltered, can be used for this. We therefore thermally evaporate 1 nm Al as seed layer on the top hBN and use atomic layer deposition (ALD, Savannah) to deposit 20 nm Al$_2$O$_3$ as additional gate dielectric and spacer between channel and gate electrodes (Fig.\ref{fig:fabrication}f).

To ensure alignment between LMH and gate SLG, we transfer a continuous film of SLG, grown by chemical vapor deposition (CVD) on Cu following the process described in Ref. \cite{Li2009Large}, using a PMMA support membrane \cite{Bonaccorso2012Production} on the SLG/Cu substrate. We etch Cu in ammonium persulfate, transfer the PMMA/SLG stack onto the photonic chip (Fig.\ref{fig:fabrication}g), and remove the PMMA by immersion in acetone. We then use two additional EBL steps, O$_2$ plasma etching, and electron beam evaporation to define the SLG split-gate geometry (Fig.\ref{fig:fabrication}h) and fabricate metal contacts to these gates (Fig.\ref{fig:fabrication}i). Finally, we perform optical lithography on a laser writer (MicroTECH LW405) and wet etching in HF to get access to drain and source contact pads.

A microscope image of the hBN/SLG/hBN LMH on the WG is in Fig.\ref{fig:material}a. We perform Raman spectroscopy (Renishaw inVia at 514.5 nm, power $<$0.5 mW) and atomic force microscopy (AFM, Bruker Dimension Icon) to monitor the SLG quality and choose a suitable device position. A typical Raman spectrum before further processing of the stack is in Fig.\ref{fig:material}b. The position of the combined hBN E$_\text{2g}$ peaks \cite{Reich2005resonant} from top and bottom flakes is Pos(E2g) $\sim$1366 cm$^{-1}$ with full-width half maximum, FWHM(E2g) $\sim$9.5 cm$^{-1}$, as expected considering the top flake is bulk and the planar domain size in these MC-produced hBN crystals is only limited by the flake size \cite{Nemanich1981light,purdie2018cleaning}. The position and full width at half maximum of the 2D and G peaks are Pos(2D) $\sim$2693 cm$^{-1}$, FWHM(2D) $\sim$ 18 cm$^{-1}$, Pos(G) $\sim$1583 cm$^{-1}$ and FWHM(G) $\sim$ 14 cm$^{-1}$, confirming the presence of SLG and low $n$ ($<$10$^{12}$ cm$^{-2}$) \cite{Das2008Monitoring}. This is confirmed by the area (A(2D)/A(G) $\sim$ 10.7) and intensity (I(2D)/I(G) $\sim7.6$) ratios, which indicate a Fermi level $E_\text{F}<100$ meV  \cite{Das2008Monitoring,Basko2009Electron,Ferrari2013Raman}.

The AFM scan of the overlap region between LMH and micro-ring in Fig.\ref{fig:material}c shows blister-free SLG/hBN interfaces, confirming a successful cleaning \cite{purdie2018cleaning}, apart from a bubble trapped in a cladding trench above the WG. The Raman FWHM(2D) map in Fig.\ref{fig:material}d, taken from a 20 $\mu$m $\times$ 30 $\mu$m in the centre of the LMH, shows a region with homogeneous (spread $<$1 cm$^{-1}$) and narrow ($\leq$18 cm$^{-1}$) FWHM(2D) and spots of increased ($>$21 cm$^{-1}$) FWHM(2D) that coincide with the blister position as revealed by AFM. Based on these findings, we then select the channel position (marked red in Figs.\ref{fig:material}a,c,d) to be in a blister-free region. The final device has an active area $L\times W\sim2.5\times 6$ $\mu$m$^{2}$. We use $L\sim$ 2.5 $\mu$m, which of the order of twice the cooling length $L_\text{cooling}$ in SLG ( $\sim1$ $\mu$m \cite{Ma2014Competing,Shiue2015High}, related to electron thermal conductivity $\kappa_e$ (see Methods) and interfacial heat conductivity $\Gamma\sim$ 0.5-5 MWm$^{-2}$K$^{-1}$  \cite{freitag2013increased,tielrooij2018out} via $L_\text{cooling}=\sqrt{\kappa_e/\Gamma}$ \cite{Song2011Hot}), to fully exploit the $T_e$ profile with expected maximum at $L/2$ \cite{Schuler2016Controlled,Muench2019Waveguide} at the WG centre.

\begin{figure*}
\centerline{\includegraphics[width=130mm]{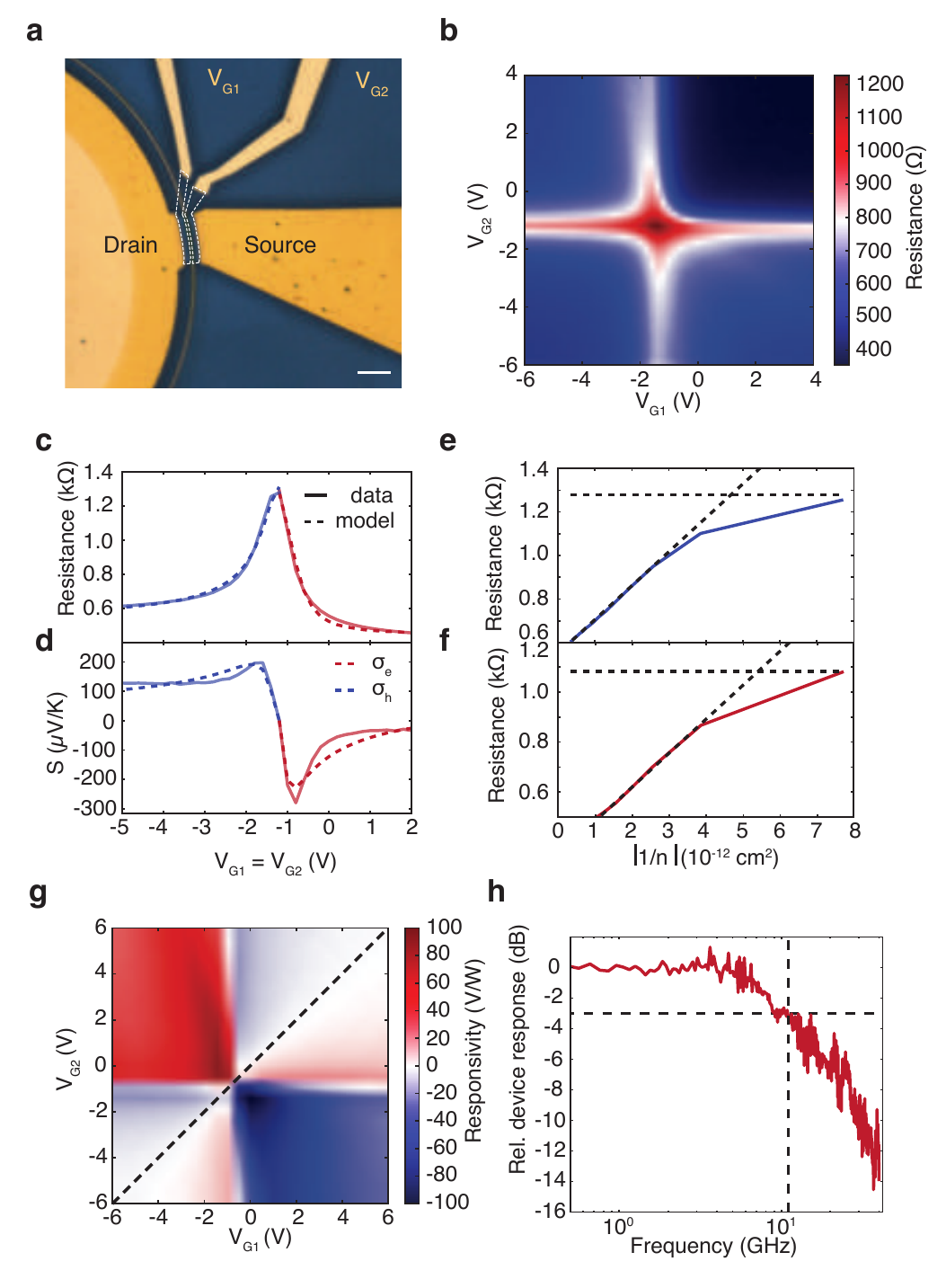}}
\caption{(a) Microscope image of device with top gate CVD SLG electrodes. Scale bar, 5 $\mu$m. (b) Resistance map demonstrating independent tunability of charge carrier concentration in the SLG channel via $V_\text{G1}$ and $V_\text{G2}$. (c) Electrical characterization at homogeneous channel doping (solid lines, measured data; dashed lines, model). (d) Calculated $S$ based on the electrical data in (c). (e,f) Resistance vs. inverse carrier concentration for (e) electron and (f) hole doping.(g) Photoresponse at zero bias on resonance ($\lambda$ = 1555.87 nm). (h) Frequency response. The 3-dB cutoff frequency, marked by intersecting dashed lines, is $\sim$12 GHz.}
\label{fig:performance}
\end{figure*}

An optical image of the final device is shown in Fig.\ref{fig:performance}a. We first verify gate tunability of the SLG channel by measuring the drain-source current ($I_\text{DS}$) at a fixed drain-source voltage ($V_\text{DS}$) while varying the two gate-voltages. The resulting resistance ($R$) map (Fig.\ref{fig:performance}b) shows a cross pattern, which confirms that four junction constellations (p-n, n-p, n-n, p-p) can be generated in the channel \cite{Schuler2016Controlled}. In order to the extract contact resistance $R_\text{contact}$ and $\mu$ from the GPD directly (rather than a four-probe reference structure made from a second LMH), then used to estimate $S$, we use the measured transfer curve at homogenous channel doping in Fig.\ref{fig:performance}c and plot $R$ as function of the inverse carrier concentration (1/$n$) for electron (Fig.\ref{fig:performance}e) and hole (Fig.\ref{fig:performance}f) doping.  By fitting the linear part (as 1/$n \rightarrow 0$) of these plots, as detailed in Ref. \cite{zhang2012direct}, $R_\text{contact}$ can be obtained from the intersection of the fit curve and y-axis ($R$), while the residual carrier concentration $n_0$ is found from intersection between the fit curve and a horizontal line through the maximum $R$. Using these values, we then model $R$ as for Ref. \cite{Schuler2016Controlled}: $R_\text{total}$=$R_\text{contact}$+$\frac{L}{W}\frac{1}{e\mu n}$, with $n=\sqrt{n_0^2+\left[C_\text{ox}/e\left(V_\text{G}-V_\text{CNP}\right)\right]^2}$, where $R_\text{contact}$ includes the contacts and the contribution from the ungated region, $V_\text{CNP}$ is the gate voltage corresponding to the charge neutrality point (CNP, $E_\text{F}$ = 0 meV), $C_\text{ox}$ is the gate capacitance, and $\mu$ is used as a fit parameter. The original data (solid line) and the model (dashed line) are compared in Fig.\ref{fig:performance}c. We get $R_\text{contact}\sim400$ $\Omega$ and $\sim530$ $\Omega$, as well as $\mu_\text{e}\sim17700$ $cm^{2}$/Vs and $\mu_\text{h}\sim11800$ $cm^{2}$/Vs for electrons (red lines) and holes (blue lines), respectively.

For optical characterisation, we first couple modulated light (ON-OFF) with a duty cycle of 50\% from a tuneable laser source (Agilent 81680A) into the bus WG using an optical single-mode fibre. While varying the potential at the two gate electrodes ($V_\text{G1}$, $V_\text{G2}$), the photoresponse is recorded using a lock-in amplifier. Fig.\ref{fig:performance}g shows a photoresponsivity map measured on resonance at 1555.87 nm, from which we extract a maximum $R_\text{[V/W]}\sim90$ V/W. The six-fold pattern, with the highest photoresponse for bipolar (p-n, n-p) junctions and a sign-change across the diagonal ($V_\text{G1}$ = $V_\text{G2}$) for unipolar (n-n, p-p) junctions in the SLG channel, confirms that the PTE effect dominates the conversion of photons into electrical signal \cite{gabor2011hot,Song2011Hot}.

To determine the BW, we modulate CW light at 1555.87 nm from the same source using a commercial (Thorlabs LN05S-FC) intensity modulator (lithium niobite, $f_\text{3dB}$ = 40 GHz) and couple it into the device. While tuning the modulation frequency of the external modulator, we monitor the GPD response with an electrical spectrum analyzer (Agilent PSX N9030A) while the gate bias ($V_\text{G1}$ = -0.5V, $V_\text{G2}$ = -2.1V) is set at an operating point where $R_\text{[V/W]}$ is largest. This gives a 3 dB bandwidth $\sim$12 GHz, Fig.\ref{fig:performance}h, as expected from the design of the passive photonic structure. In optical transmission systems that employ the non-return-to-zero (NRZ) format (i.e. two non-zero voltage levels used to represent digital ones and zero \cite{Kaminow2013Optical}) for signal encoding, the BW of the PD employed in the receiver corresponds to $\sim60-70\%$ of the bit rate (bits processed per second) the optical link can support \cite{sackinger2005broadband}. Our GPD is thus capable of detecting data rates up to $\sim\frac{f_\text{3dB}}{0.6}\sim$ 20 Gbit/s, confirming its suitability for applications in data centre optical interconnects \cite{cheng2018recent}.
\begin{figure*}
\centerline{\includegraphics[width=180mm]{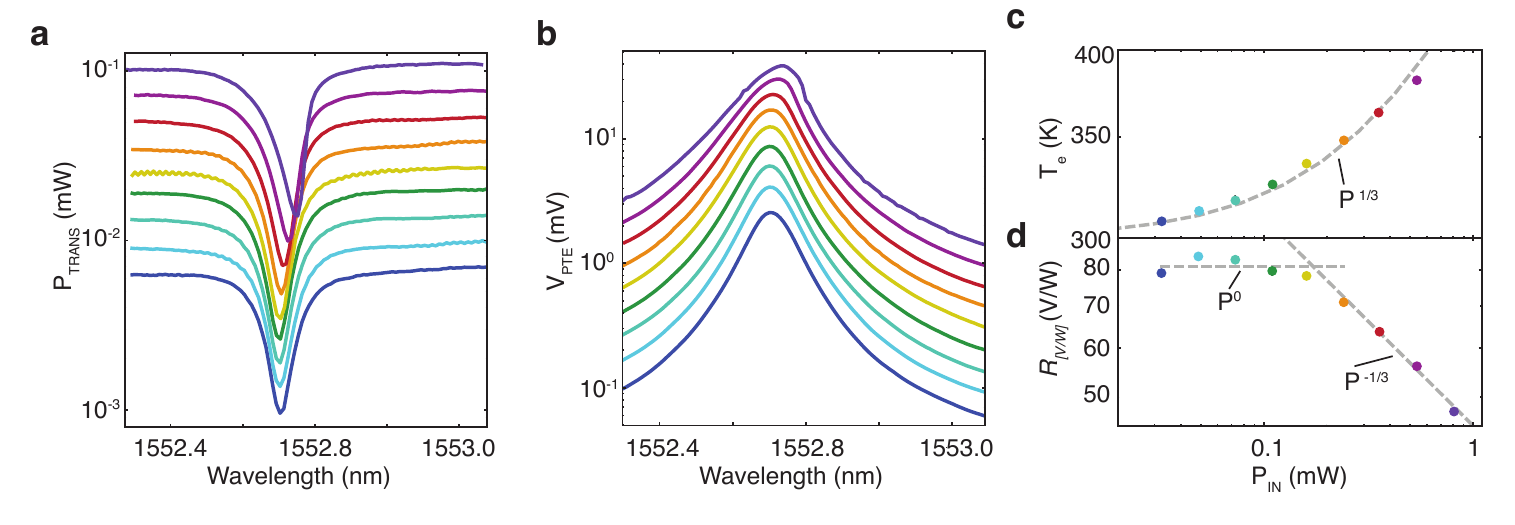}}
\caption{(a) Transmitted power for various $P_\text{in}$. (b) Photovoltage for a fixed gate voltage combination ($V_\text{G1}=1$ V, $V_\text{G2}=-1$ V) measured for various input powers (same color code as for the transmission in (a)). (c) $T_e$ calculated from the photoresponse in (b) and S in Fig.\ref{fig:performance}(d). (d) Power-dependent $R_\text{[V/W]}$.}
\label{fig:dependencies}
\end{figure*}

Figs.\ref{fig:dependencies}a,b plot the wavelength dependence of optical transmission and photovoltage for $V_\text{G1}$ = 1 V, $V_\text{G2}$ = -1 V and various $P_\text{in}$. The shift of the resonance and its asymmetry are attributed to the power dependent change of the effective refractive index of the Si WG through a thermo-optic effect \cite{bogaerts2012silicon,Fainman2013silicon}. Extracting the power-dependent maxima in $V_\text{PTE}$ allows us to estimate $T_e$ at resonance, and plot the $R_\text{[V/W]}$ power-dependence. 

The junction carrier temperature $T_\text{e,j}$ can be written as (see Methods):
\begin{equation}\label{eq:junctionCarrierT}
T_\text{e,j} = \sqrt{2\bigg|\frac{V_\text{PTE}}{\zeta_{1}-\zeta_{2}}\bigg|+T^{2}_{0}}
\end{equation}
where $T_0$ = 294 K is the room temperature and $\zeta_{1,2}=\frac{\pi^2 k_\text{B}^2}{3e\sigma}\frac{\mathrm{d}\sigma}{\mathrm{d}\epsilon}$ at $\epsilon=E_F$ in analogy to Eq.\ref{eq:Mott}. In order to estimate $\zeta_{1,2}$, following the same method used to determine $S$ in Fig.\ref{fig:performance}d, we use $R_\text{contact}$ and $\mu$ as obtained from the electrical measurements at homogeneous channel doping. The resulting $T_e$ extracted for different $P_\text{in}$ is shown in Fig.\ref{fig:dependencies}c. The carriers reach $T_e\sim$ 400 K for $P_\text{in}\sim$ 0.6 mW, due to the cavity-enhanced light-matter interaction. The $T_e$ power-dependence can be fitted by the heat equation \cite{Song2011Hot,Ma2014Competing}, neglecting diffusive cooling through the contacts (see Methods) as:
\begin{equation} \label{eq:Tbehaviour}
T_\text{e,j} = \left(\beta P_\text{IN} + T^{\delta}_0\right)^\frac{1}{\delta}
\end{equation}
with $\delta\sim3$ \cite{graham2013photocurrent} and $\beta$ a fitting parameter.

The associated, power-dependent $R_\text{[V/W]}$ is presented in Fig.\ref{fig:dependencies}d. For small $P_\text{in}$ ($<$0.2 mW), we obtain a constant $R_\text{[V/W]}$. For $P_\text{in}>$0.2 mW we get sublinear scaling between $V_\text{PTE}$ and $P_\text{in}$, resulting in a drop of $R_\text{[V/W]}$ according to $R_\text{[V/W]}\propto P_\text{in}^{-1/3}$ due to the $T_e$ dependence of the electronic heat capacity \cite{graham2013photocurrent,tielrooij2015generation}.

In order to assess our GPD performance against non-SLG based ones, where typically a photocurrent $I_\text{ph}$ is generated \cite{chrostowski2015silicon}, we compare two representative optical receiver implementations: a 'conventional' (i.e. non-SLG) system based on a wafer-scale, commercial, high-speed ($>$50 GHz) Ge photodiode with a (current) responsivity $R_\text{[A/W]}=I_\text{ph}/P_\text{in}\sim0.5$ A/W \cite{absil2015silicon}, and a receiver based on our GPD. In both systems the same voltage amplifier is employed to obtain a $>$200 mV output voltage swing ($V_\text{OUT}$), as required for driving the subsequent clock and data recovery CMOS circuit. For simplicity we assume that the amplifier represents a capacitive load.

Our GPDs are based on the PTE effect where an electromotive force directly provides a voltage, rather than a current \cite{Koppens2014Photodetectors,tielrooij2015generation}. In case of Ge, an additional TIA is needed to convert the photocurrent into a voltage for further signal processing \cite{okamoto2016chip}. In the TIA we consider a feedback resistor $R_\text{F}$ = (90V/W)/(0.5 A/W) = 180 $\Omega$, which assures the same $V_\text{OUT}$ for same optical input power in both cases. Neglecting any noise other than thermal noise produced by $R_\text{F}$, we estimate for the conventional receiver a lower limit for the sensitivity  \mbox{$\bar{P}_\text{sens}=Q i_\text{n}/R_\text{[A/W]} = 12.6$ $\mu$W $\sim$ -19 dBm} at a bit-error-rate (i.e. probabilty of false identification of a bit by the receiver decision circuit  \cite{agrawal2012fiber}) BER = $10^{-9}$. Here we calculated the thermal noise current as \mbox{$i_\text{n}$ =$\sqrt{4 k_\text{B} \cdot T \cdot BW/R_\text{F}}$} with BW = 12 GHz, as in our GPDs, and a $Q$ factor (i.e. required signal-to-noise ratio to get a specific BER \cite{bergano1993margin}) $\sim$6 from \cite{agrawal2012fiber} \mbox{BER = $\frac{1}{2}\text{erfc}(Q/\sqrt{2})$}. For our GPDs, we estimate \mbox{$\bar{P}_\text{sens}$ = $Q v_\text{n}/R_\text{[V/W]}\sim$ -16 dBm} for same BER and $BW$, where \mbox{$v_\text{n}$ = $\sqrt{4 k_\text{B}\cdot T \cdot BW \cdot R_\text{G}}$} and $R_\text{G}\sim800$ $\Omega$ is the total device resistance. Thus, the $\bar{P}_\text{sens}$ of our GPD-based receiver is on par with mature semiconductor technology and could be further improved by reducing $R_\text{contact}$, which dominates the total device resistance, thus being the primary source of thermal noise. The natural generation of a voltage makes the need for a TIA obsolete, with a reduction of the energy-per-bit cost and system foot-print.
\section{Conclusions}
We reported photo-thermoelectric GPDs, integrated on Si microring resonators. By tuning the SLG coverage on top of the resonator, we optimised the round-trip propagation losses inside the cavity to achieve critical coupling, achieving $>90$\% light absorption in $\sim6$ $\mu$m SLG. The resulting carrier heating on such compact lengths enables high peak $T_e\sim$ 400 K in the GPDs. In combination with high ($>10^4\text{ cm}^2/\text{Vs}$) mobility, achieved by encapsulating the SLG channel in hBN, this allowed us to achieved $R_\text{[V/W]}\sim$ 90 V/W. Our bias-free and Johnson-noise limited GPDs with voltage output are a more power-efficient alternative to state of the art commercial PDs used in optical interconnects. Hot-carrier effects in SLG can be used for receiver architectures where current-to-voltage conversion, traditionally performed by transimpedance amplifiers, can be bypassed for a reduction in energy-per-bit cost and system foot-print.
\section{Methods}
The junction $T_e$ can be extracted from the power dependent photovoltage measurements. Rewriting Eq.\ref{eq:Mott} as \mbox{$S=\zeta T_\text{e}$}, where $\zeta=-(\pi^2 k_\text{B}^2)/3e\sigma\mathrm{d}\sigma/\mathrm{d}\epsilon$ at $\epsilon=E_F$, and using this in Eq.\ref{eq:V-PTE}, we get $V_\text{PTE}=\int S(x) \nabla T_\text{e}(x)\mathrm{d}x = \int \zeta(x) T_\text{e}(x) \nabla T_\text{e}(x)\mathrm{d}x$. Integration over the p-n-junction gives $V_\text{PTE}=\zeta_1\int^0_{-\frac{L}{2}}T_\text{e}(x)\frac{\mathrm{d}T_\text{e}}{\mathrm{d}x}\mathrm{d}x+\zeta_2\int^{\frac{L}{2}}_0 T_\text{e}(x)\frac{\mathrm{d}T_\text{e}}{\mathrm{d}x}\mathrm{d}x=\frac{\zeta_1-\zeta_2}{2}\left(T^2_\text{e,j}-T^2_0\right)$. From this, we extract the junction $T_e$ from the measured photovoltage as $T_\text{e,j}$=$\sqrt{2\bigg|\frac{V_\text{PTE}}{\zeta_{1}-\zeta_{2}}\bigg|+T^{2}_{0}}$. In the most general case, the $T_e$ profile can be calculated from the heat equation \cite{Song2011Hot,Ma2014Competing}, including diffusive cooling through contacts and cooling through the phonon bath: $\frac{\mathrm{d}q}{\mathrm{d}x}=-\frac{\mathrm{d}}{\mathrm{d}x}\left(\kappa_\text{e}\frac{\mathrm{d}}{\mathrm{d}x}T_\text{e}\right)+\frac{\kappa_\text{e}}{L_\text{cooling}^2}\left(T^{\delta}_\text{e}-T^{\delta}_\text{0}\right)$, where $\frac{\mathrm{d}q}{\mathrm{d}x}$ describes the heating of the system. The electronic thermal conductivity is given by $\kappa_\text{e}=\sigma \mathcal{L}_0 T_\text{e}$ \cite{kittel1996introduction}, with Lorenz number $\mathcal{L}_0$, while a $T_e^{\delta}$ dependence with $\delta\sim3$ is characteristic for graphene  \cite{Ma2014Competing,Song2012Disorder}. Assuming that the distance from the heat source to the heat sink larger than the cooling length ($L_\text{channel}>2 L_\text{cooling}$), the heat equation becomes $\frac{\mathrm{d}q}{\mathrm{d}x}=\frac{\kappa_\text{e}}{L_\text{cooling}^2}\left(T^{\delta}_\text{e}-T^{\delta}_\text{0}\right)$, with $\frac{\mathrm{d}q}{\mathrm{d}x}=\alpha P_\text{in}/A$, with $\alpha P_\text{in}$ the fraction of the absorbed power, and $A$ the heated area. We thus get $T_\text{e,j}=\left(\beta P_\text{in}+T^{\delta}_0\right)^\frac{1}{\delta}$, with $\beta$ a fit parameter.
\section{Acknowledgments}
We thank Stefan Wachter, Daniel Schall, and Andrea Tomadin for useful discussions. We acknowledge funding from the EU Graphene Flagship, ERC grant Hetero2D, DSTL, EPSRC grants EP/L016087/1, EP/K01711X/1, EP/K017144/1, EP/N010345/1
\section{Author contributions}
S.S. and T.M. conceived the device. S.S. and J.E.M. fabricated the devices and performed the experiments. S.S., J.E.M., T.M., and A.C.F. analysed and discussed the data. J.E.M. and A.R. built the measurement set-up. O.B. prepared CVD SLG. T.T. and K.W. prepared hBN. D.V.T. and V.S. designed and provided the SOI waveguide substrates. T.M., A.C.F., I.G., and M.R. supervised the work. S.S., J.E.M., T.M., and A.C.F. wrote the paper, with input from all authors.

\end{document}